# World-wide Evidence for Gender Difference in Sociality

Tamas David-Barrett

Email: tamas.david-barrett@trinity.ox.ac.uk

Address: Trinity College, Broad Street, Oxford, OX1 3BH, UK

Web: www.tamasdavidbarrett.com

**Abstract**

One of the most contested questions about human behaviour is whether there are inherent sex or gender differences in the formation and maintenance of social bonds. On one hand, female and male brains are structurally almost identical, and while there are sex differences in the endocrine system, these are small, while much of gendered identity and behaviour is learned. On the other hand, sex differences in some aspects of social behaviour have deep evolutionary roots, and are widely present in non-human animals. This observational study recorded the frequency of same-aged, adult human groups appearing in public spaces through 2636 hours, recording group formation by 1.2mn people via 170 research assistants in 46 countries across the world. The results show (a) a significant sex-gender difference in same-sex-same-age frequency, in that ~50% more female-female than male-male pairs are observed in public spaces globally, and (b) that despite regional variation, the patterns holds up in every global region. This is the first study of sex-gender difference in dyadic social behaviour across the world on this scale, and the first global study that uses direct rather than internet-based observations.

# Introduction

One of the most intriguing and at the same time most contested questions about human sociality is whether there are sex or gender differences in the way we form and maintain our bonds.

On one hand, the neurological differences between female and male brains are small, and almost entirely due to these brains having to 'run' somewhat different bodies [1-3]. There is no sex difference in general intelligence [4-7], although there is a sex difference in the location where the cognition associated with general intelligence occurs in the brain [8-10], as well as in the inter-individual variability of brain structure [11, 12]. At the same time, gender is at least to some extent a learned identity and behaviour [13-16], and goes beyond a strict sex-to-gender binary assignment in every culture in which this question has been studied [17-20].

On the other hand, there is a well-established sex difference in the endocrine system [21-23], which has deep evolutionary roots [24-26]. This difference explains the behavioural sex differences in non-human animals [27-31], and a host of social behaviours in humans [32-39].

In other words, the female and male neurone 'computers' are almost the same, the emotions are only a little different, and we learn much of our gendered behaviour anyway.

The fact that human social behaviour is rooted in evolution, while it is also flexible, would lead us to believe that social behaviours with deeper-older origins and more fixed evolutionary functions would show more sex, and thus gender, difference than newer and less fixed ones [40-43]. We would expect that behaviours around reproduction, such as, pair bond, sexual behaviour, and parenting, would show stronger sex and gender difference than in friendship.

In fact, the modern phenomenon of friendship-based societies is fairly new in our evolutionary history [44, 45], probably brought on by demographic processes, such as urbanisation and fertility transition [46, 47].

Although sex-gender differences in friendship had been observed in many cultures and settings [48-56], the first comprehensive global study took place only less than a decade ago [57]. This study of ~112,000 Facebook profile pictures showed a striking gender difference in both the propensity to create dyadic connections, and the affinity towards large groups, a phenomenon that held across all global regions. The finding was in line with the general pattern that women tend to invest more into intensively maintaining social relationships than men do [58-61].

One of the possible drawbacks of the Facebook study was that it was based on people appearing together on a Facebook profile picture, and hence was subject to two possible biases. First, only internet users with Facebook account could be sampled, raising the possibility of selection bias [62-64]. Second, people choose their profile pictures from a range of photos, and thus this selection possibly suffered from vanity- or identity-type selection bias [65-67].

This current paper's methodology alleviates both of these concerns. A way around these forms of selection bias is to observe people in their natural habitat without their awareness to the observation, i.e., to use the approach of human ethology [68]. Natural human habitat today is in cities, towns, and villages, and the space where ethical observation can be made without detection is in public spaces. This paper is the first of that uses this new dataset collected by 170 research assistants in 46 countries around the world.

This paper will test the following hypotheses:

Hypothesis 1: the frequency of female-female dyads is higher than the frequency of male -male dyads. That is, does the frequency differential found in the Facebook study repeat in public spaces?

Hypothesis 2: the frequency pattern of gender difference in sociality is present in all parts of the world. That is, does the universality of gender difference in same-sex two person groups hold across human societies, implying that the underlying preference is robust to cultural variation?

# Methods

I recruited research assistants via national psychological associations and psychology departments of universities around the world. Approximately 500 applications came in, of which I conducted interviews with approximately 350, of which approximately 250 signed up for the project. Almost all RAs were either psychology students or practicing clinical psychologists. All RAs were blind to the research hypothesis.

Altogether 171 research assistants submitted data, from observations in 46 countries on six continents. I excluded the data from one RA who (a) submitted tally sheets of 20 hour-long daily observations on consecutive days, (b) with 10,000+ observations per tally sheets, which is physically impossible, and (c) did not submit the hard copy of these tally sheets (which most RAs did). I ran extensive statistical tests to see if other RAs might have made up the submitted data, and I found no reason that would have justified further exclusion. This left data from 170 RAs in the dataset.

The research assistants were instructed, both verbally on a skype call and in writing (see Supplementary Material 1), to find a coder position in a public space, from which to observe people going about their daily lives, and record their observations on physical copies of a tally sheet (see Supplementary Material 2). The tally sheet was designed to be as similar to the 'profile picture' frequency study's tally sheet [57] as possible. The RAs received detailed instructions about the coding categories (see Supplementary Material 3).

The RAs submitted altogether 2731 tally sheets based on 2636 hours of observing more than 1,192,000 people. Most of the observations took place during 2014, and all finished by early 2015.

The data cleaning took substantial amount of time: most problems were due to the regional and language variation of MS Excel, the software in which most RAs reported their results, as well as due to a few RAs having somewhat redesigned the report sheet.

I analysed the data using Wolfram Mathematica.

**Research ethics**

Every research assistant of the project participated in and received a certificate from the PHRP online research ethics course. The research design was approved by the Oxford University Medical Science Ethics committee, approval number MSD-IDREC-C1-2014-122.

# Results

In line with the code book instructions, the gender of the observed people was registered on the tally sheet only in the case when a same age adult was either alone, or was part of same-aged group and, in the judgement of the coder, could be categorised either as female or male. Thus, out of the ~1,192,000 observed people, the dataset contains the observed-gender of 982,424 people, all adults, all appearing either alone or as part of a same-aged group. Out of this total, 479,921 were categorised as women, and 502,503 as men.

The focus of this paper is the frequency difference between the 2F and 2M categories. There were 51,033 two-women groups, and 41,416 two-men groups. Thus, in this dataset there are 23.2% more 2F pairings than 2M pairings, and the 2F frequency is higher than the 2M at $p<10^{-13}$ (Fig. 1).

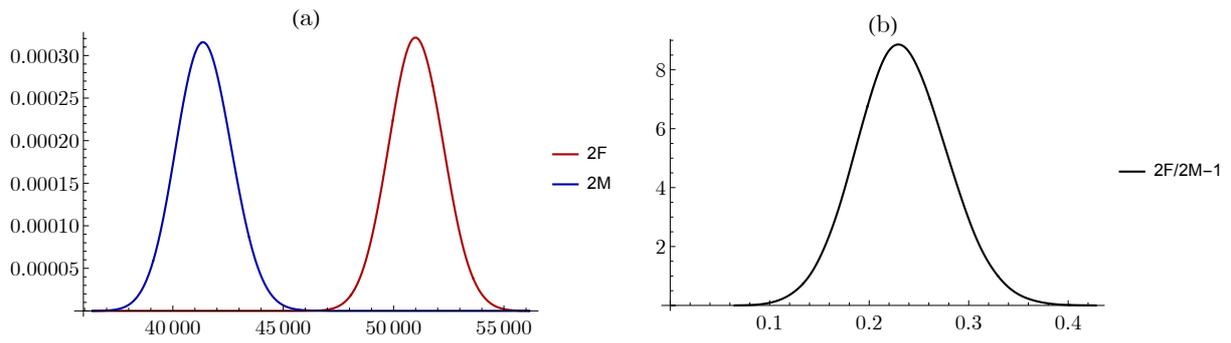

Fig. 1. Bootstrap histograms of the frequency of two-women (2F), and two-men (2M) groups in the dataset. Panel (a): bootstrap histograms, panel (b) histogram of the 2F/2M-1 ratio. (For all three histograms: 50,000 re-samplings of 2735 length.)

Notice that although the 2F/2M ratio in the current study is much lower than in the 'profile pictures' study: 23.2% compared to 50.1%, the comparison is not entirely fair. The 'profile pictures' study had somewhat more women, and the current 'public spaces' study has somewhat more men. To correct for this, I used the approach of the 'profile pictures' paper [57], by employing single person frequencies as inverse weight to the 2F/2M ratio, yielding the adjusted ratio of (2F/1F)/(2M/1M)-1, which is 41.8% in the 'profile pictures' paper and is 49.7% in the current 'public spaces' study. These are not significantly different from each other (Fig. 2).

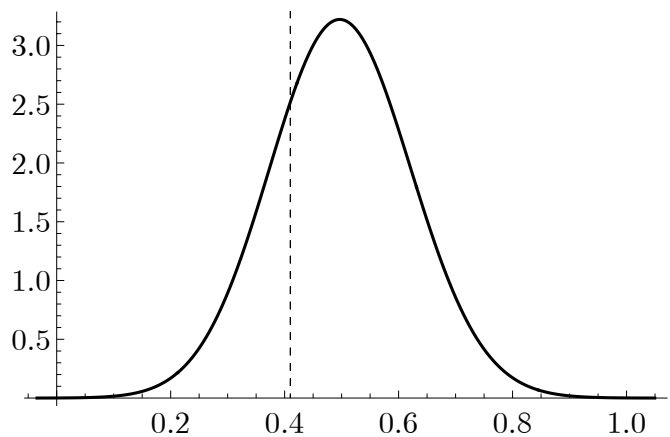

Fig. 2. Comparing the bootstrap histogram of the 2F/1F/2M/1M-1 to the same measure in the 'profile pictures' study (dashed line). The two means are not significantly different.

Given geographical distribution of the 46 countries in the dataset, it is possible to define four global regions (SM 4). The adjusted ratio is significantly above zero in every region (Fig. 3).

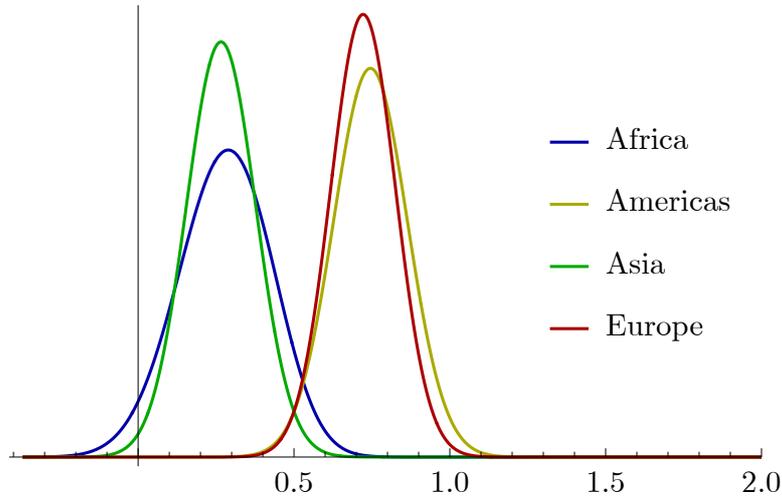

Fig. 3. Variation among global regions. The histograms are bootstraps, from which the p-values are calculated (50,000 repeats of the same lengths as telly sheets from that region). For country categorisations, see SM4.

Thus, findings support both hypotheses 1 and 2: there is a significant gender difference in the frequency of same-sex two-person groups, which holds for every global region, bar one region in a single measure (Table 1). This result is robust to coder and within region country variation, and transformation using observation hours as weights (SM5-6).

**Table 1. Regional statistics**

|  | **Africa** | **Americas** | **Asia** | **Europe** |
|---|---|---|---|---|
| Number of people observed | 77,954 | 227,292 | 228,126 | 660,735 |
| of which |  |  |  |  |
|    women | 32,216 | 87,657 | 84,812 | 275,236 |
|    men | 34,536 | 96,453 | 114,898 | 256,616 |
|    gender not assigned | 11,202 | 43,182 | 28,416 | 128,883 |
| Hours observed | 218 | 480 | 557 | 1,381 |
| Adjusted 2F2M ratio | 0.27 | 0.74 | 0.27 | 0.72 |
| Number of tally sheets | 198 | 540 | 557 | 1,436 |
| Tally-sheet-bootstrapped p | 1E-90 | 0 | 0 | 0 |
| Bootstrapped distribution p | 0 | 0 | 0 | 0 |
| Number of coders | 18 | 35 | 32 | 98 |
| Mean of coder means | 0.49 | 1.07 | 0.78 | 0.95 |
| Median of coder means | 0.41 | 0.81 | 0.63 | 0.75 |
| Coder-bootstrapped p | 0.002 | 1E-06 | 7E-05 | 5E-17 |
| | | | | |
| Number of countries | 5 | 4 | 12 | 25 |
| Mean of country means | 0.26 | 0.86 | 0.77 | 0.72 |
| Median of country means | 0.29 | 0.8 | 0.71 | 0.71 |
| Country-bootstrapped p | 0.233 | 0.009 | 0.001 | 2E-08 |

For the region definitions, see SM4.

'Countries' refers to the number of countries in which observations were made.

'Coders' refers to the number of coders who made observations in this region. (Note that some coders made observations in more than one region.)

'Hours' refers to the number of observation hours.

'Women', 'men' refer to the number of women and men in who are registered in the tally sheet with gender.

"NA' refers to the number of people who are registered in the tally sheet without their gender being recorded. Note that this number is a minimum estimate as any of the NH, NP, CB, MP, and CTG categories can contain more than one person, and some of the categories (e.g., NP) likely often included families of more size more than 2. (For code definitions and instructions, see SM1-3 for.)

'Total' refers to the total number of people observed. (This is a minimum estimate, see above.)

'Ratio' refers to the 2F/1F/2M/1M-1 ratio.

'p-value' is calculated from the bootstrap histogram (50,000 repeats of length same as the number of tally sheets or coders or countries, respectively, in the region's dataset).

(NB. Given that there is substantial regional variation, the finding that the previous 'profile picture' and the current 'public spaces' datasets yielded almost the same headline measures is likely accidental.)

# Discussion

The mainline result presented in this paper is consistent with the hypothesis that women prefer dyadic same-sex friendships more than men do [50, 69-71], while the geographical pattern found is consistent with the hypothesis that this behaviour is universal in humans.

These results are consistent with two, opposing causal explanations. On one hand, it has been suggested that the universality of the pattern supports the hypothesis that this gender difference has an inherited component and is likely to be an evolved strategy to cope with ancestral patrilocality [57, 72]. This hypothesis is further supported by the observation that both bonobos (*Pan paniscus*) [73-75] and chimpanzees (*Pan troglodyte*) [76-78] practice patrilocality, and both species have strong non-kin female dyadic bonds [31, 79, 80] suggesting at the minimum a correlation between the two traits. The literature tends to favour the view that patrilocality was the human ancestral state, too [81-85] (although with disagreements [86]), suggesting that the evolutionary logic to how sister-like friendship can be adaptive in patrilocal societies worked in our species, as well.

On the other hand, the universality of a behaviour makes an evolutionary origin likely, but not guaranteed. There are shared traits of the societies in which this paper's observations were made which has to do with our current human societies organising themselves, and thus a shared cultural or institutional logic, and not an evolutionary one that may have been driving the pattern observed in this paper. For instance, every one of the societies in the dataset have a patriarchal history, and every one of them is dominated by a market economy to which access is gender biased at least to some extent.

This dataset's size makes it possible to test a range of causal hypotheses, which will be the subject of future work.

# Supplementary Materials

**Supplementary Material. 1. Coding instructions**

# Instructions

**For Researchers Participating In The Global Data Collection Of
The Human Groups In Public Spaces Project**

First of all, thank you for your participation in *The Human Groups In Public Spaces Project*. Welcome in the team.

This document contains the instructions for the researchers involved in the data collection in our global effort to map human groups in public spaces.

Your task is to observe human groups in public spaces, and record their sizes and gender composition in tally sheets ('*public spaces tally sheet.docx*'). Please print the tally sheets in as many copies as observation events, and use a new tally sheet for each separate observation. Please keep these sheets. At the end of the data collection phase of the project, we will ask you to send us the actual tally sheets in either hard copy, or digital copies in email.

To make the observations, please find a suitable observation point in a public space, preferably somewhere where there is a relatively large number of people appearing or regularly crossing. For instance, a busy street, a village/town/city square, a park, a beach, a train station, a department store, an airport would be perfect, but if you have any other ideas, especially if they are local specialty, please go ahead and be inventive. The objective of the study is that we have a representative sample, and hence would wish to sample in as many different locations as possible.

Once you have your observation post (a café table in a square, a bench in the park, etc.), please start with filling in the header of the tally sheet: your name, the location (country, city, exact address if it makes sense, as well as type: e.g., park or airport or beach or what else best describes the nature of the public space you are observing people in), the date, and the start time. (When you finish the observation, please do not forget to fill in the duration as well.) It is very important that you fill these in, although at the time they might seem trivial. (In the data analysis process we will have to run specific test for each coder, type of location, etc., for which we must be able to tell which observation belongs to which category.)

When you are done with the header, you are ready for the observations. Please record the human groups and their compositions on the tally sheet. The category definitions you will find in the *'Code definitions.docx'* file. Here are a few points to remember:

1. You will notice that if the group you see involves a child (or is a child alone), or people in different generations, then the group will fall into either NP or CB category. In other words, groups that are recorded in 1F down to 5+ categories should include only same generation members of the group ALL of whom are adults. (Whether they are of different generations, or adults, is up to you to decide. We trust your judgment.)
2. Whether people are dressed up in fancy dress (MP), or what the genders of the different members of the group are, it is also up to your judgment. (If you can't tell the gender, just tally CTG.)
3. The categories 1F to 3F1M should all be filled using tallies. That is, if you see a single woman: tally 1F, if you see three men forming a group: tally 3M, if you see two women with one man: tally 2F1M.
4. For the 5+ category, please spell the group compositions out; for instance, if you see three women and two men: 3F2M; or if you see a group made up of six women: 6F.

VERY IMPORTANT: under no circumstances should you approach anybody you are observing, you should not record any personal detail beyond the tally sheet, or make any photos, videos. We are following the Oxford University ethics guidelines: our research is entirely anonymous, and we must make sure that it stays that way.

Once you have done at least a few observations (let's say, at least 3 hours in at least two chunks), please fill out the data report sheet (you will find an example in the *'public spaces data report example.csv'* file), and send them to me (tamas.david-barrett@psy.ox.ac.uk). I will check that there are no misunderstandings, misinterpretations, and then you can do the observation in bulk, and only send me the data (to tamas.david-barrett@psy.ox.ac.uk) after each 20 or so hours. Please name your report file the following way:

*'public spaces data report, YOURNAME, REPORTNUMBER.csv'*, if you prefer it is also fine if you send your data in excel format

Thank you!

PS. A word on the hypothesis. This is something that we cannot tell you at the moment, for it might unconsciously affect your coding judgment. Hence we need to keep you 'blind', for the moment. However, once the data collection phase is over, we will let you into the hypotheses without delay.

# Supplementary Material 2. Tally sheet

Coder:    Location:    Date:    Start time:    Total mins:

|   |      |   |
|---|------|---|
|   | NH   |   |
|   | NA   |   |
|   | NP   |   |
|   | CB   |   |
|   | MP   |   |
|   | CTG  |   |
|   | 1F   |   |
|   | 2F   |   |
|   | 3F   |   |
|   | 4F   |   |
|   | 1M   |   |
|   | 2M   |   |
|   | 3M   |   |
|   | 4M   |   |
|   | 1F1M |   |
|   | 1F2M |   |
|   | 2F1M |   |
|   | 1F3M |   |
|   | 2F2M |   |
|   | 3F1M |   |
|   | 5+   | Detailed code for 5+: |

Notes:

# Supplementary Material 3. Code definitions

| Code | Description |
| --- | --- |
| NH | Not human: e.g., object, landscape, monster, car, or any human appearing with an animal or large object |
| NA | [Leave this empty] |
| NP | Multiple people but not peer: e.g., mother-child, a family |
| CB | Child or baby alone |
| MP | People dressed up in a fancy dress etc. |
| CTG | Can't tell gender |
| 1F | 1 female |
| 2F | 2 females |
| 3F | 3 females |
| 4F | 4 females |
| 1M | 1 male |
| 2M | 2 males |
| 3M | 3 males |
| 4M | 4 males |
| 1F1M | 1 female + 1 male (e.g. a heterosexual couple) |
| 1F2M | 1 female + 2 males |
| 2F1M | 2 females + 1 male |
| 1F3M | 1 female + 3 males |
| 2F2M | 2 females + 2 males (e.g. two couples) |
| 3F1M | 3 females + 1 male |
| 5+ | Five or more people on the picture [please add detailed code as a list] |

**Supplementary Material 4**

Country assignments to continents:

Africa: Kenya, Namibia, South Africa, Zambia, Zimbabwe.

Asia: Australia, Cambodia, China, India, Indonesia, New Zealand, Qatar, Saudi Arabia, Singapore, Sri Lanka, Thailand, UAE.

Europe: Austria, Belgium, Croatia, Cyprus, Czech Republic, Denmark, Estonia, France, Germany, Greece, Hungary, Iceland, Ireland, Israel, Italy, Latvia, Netherlands, Portugal, Russia, Slovakia, Slovenia, Spain, Sweden, Switzerland, UK.

Americas: Argentina, Bahamas, Mexico, USA.

**Supplementary Material 5: Regional bootstrap histograms, by measure**

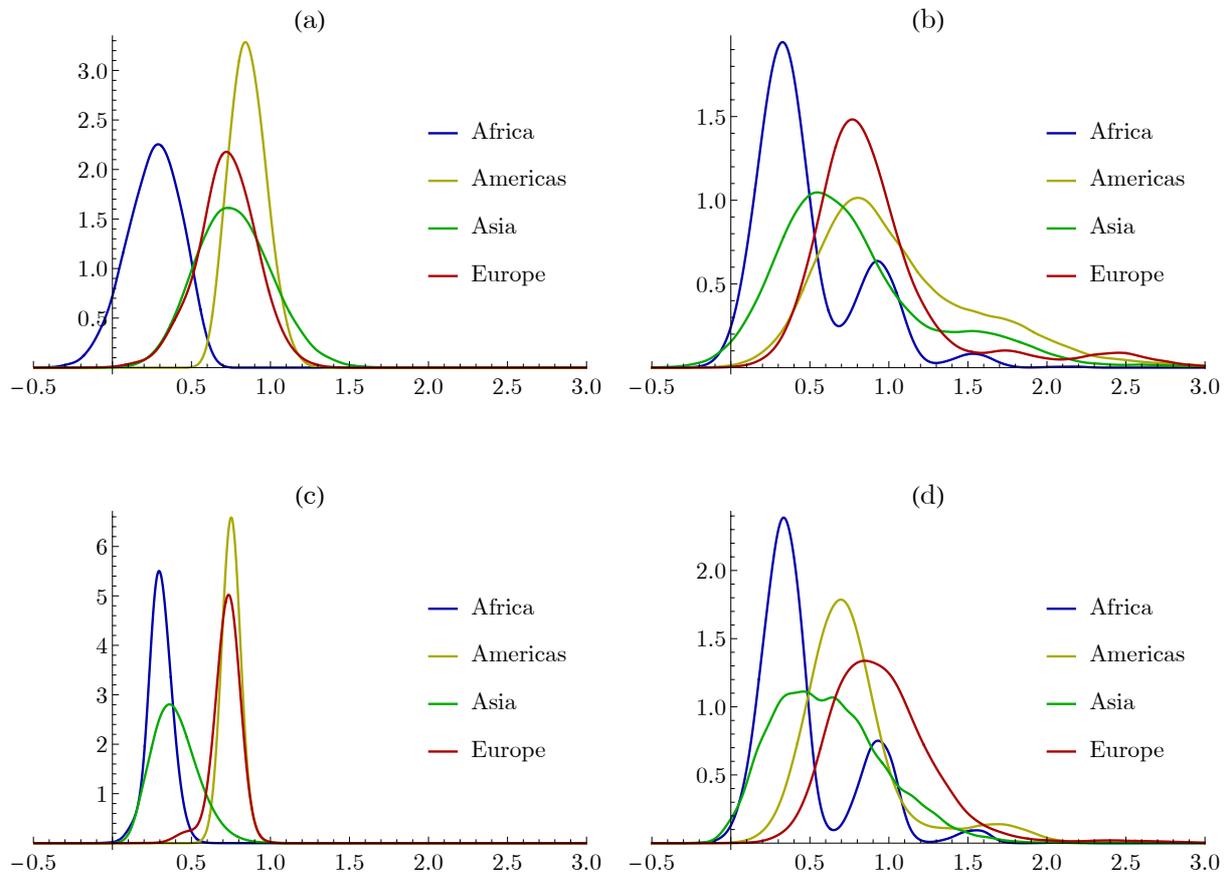

Fig. S1. Regional country- and coder-bootstrapped histograms, by measurement type. (Panel (a): bootstrapped histogram of country averages; (b): of coder average; (c): of country averages weighted by observation hours; (d): coder averages weighted by observation hours.)

# Supplementary Material 6: Regional bootstrap histograms, by region

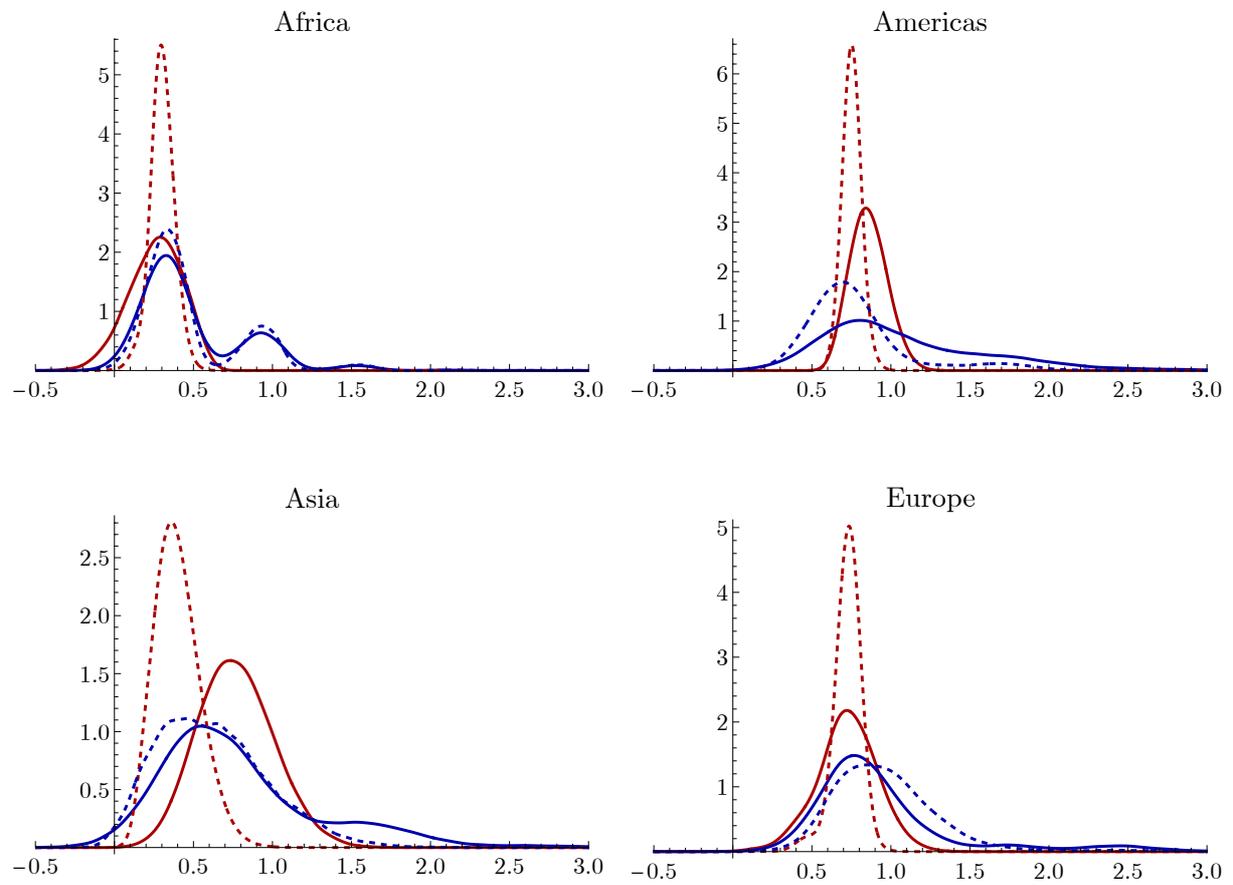

Fig. S2. Regional country- and coder-bootstrapped histograms, by region. (Red: bootstrapped histogram of country averages; dashed red: of country averages weighted by observation hours; blue: of coder average; blue dashed: coder averages weighted by observation hours.)